# Model-based Task Analysis and Large-scale Video-based Remote Evaluation Methods for Extended Reality Research

**Yalda Ghasemi**
University of Illinois at Chicago, zghase3@uic.edu
**Heejin Jeong**
University of Illinois at Chicago, heejinj@uic.edu

In this paper, we introduce two remote extended reality (XR) research methods that can overcome the limitations of lab-based controlled experiments, especially during the COVID-19 pandemic: (1) a predictive model-based task analysis and (2) a large-scale video-based remote evaluation. We used a box stacking task including three interaction modalities - two multimodal gaze-based interactions as well as a unimodal hand-based interaction which is defined as our baseline. For the first evaluation, a GOMS-based task analysis was performed by analyzing the tasks to understand human behaviors in XR and predict task execution times. For the second evaluation, an online survey was administered using a series of the first-person point of view videos where a user performs the corresponding task with three interaction modalities. A total of 118 participants were asked to compare the interaction modes based on their judgment. Two standard questionnaires were used to measure perceived workload and the usability of the modalities.

CCS CONCEPTS • Human-centered computing • Human computer interaction (HCI) • HCI design and evaluation methods

**Additional Keywords and Phrases:** Extended reality, Model-based evaluation, Remote evaluation, Gaze interaction

## 1  INTRODUCTION

Interaction modalities in extended reality (XR) have been evaluated using human subjects. The traditional evaluation has been done in the laboratory setting, but there are some issues, such as the limited number of participants, difficult recruitment process, cost-sensitivity, and risks associated with direct contacts specifically during the COVID-19 pandemic. As an alternative, a model-based evaluation technique (i.e., Goals, Operators, Methods, and Selections rules - GOMS) can be used to evaluate some aspects of a system before conducting an in-person experiment by defining the task operators and predicting the task execution time for the user interactions in XR. This method helps researchers to understand human behaviors and predict task execution times without any human-subject experiments.

Another method to evaluate the XR interactions is to perform a large-scale online survey. In the online survey, we present participants the pre-recorded video clips of user interactions and request them to respond to a series of questions from two standard questionnaires (i.e., NASA-Task Load IndeX (NASA-TLX) and System Usability Scale (SUS)).

## 2  RELATED WORK

### 2.1  Model-based Evaluation

Model-based evaluation has been used in the field of human-computer interaction (HCI) for modeling the human performance and determining how a human uses a specific system. Since experiments involving human subjects are often cost and time intensive, it is beneficial to apply these techniques beforehand as they provide low-cost usability evaluations [1]. GOMS was proposed by [2] as a family of predictive models for measuring human performance with regard to their task execution time in interactive systems. The total time to complete a task is calculated by summing the time of its relevant operators. GOMS has been investigated as a predictive tool in many research studies [3, 4]. Pointing is a primary operator of the GOMS models. Earlier studies on GOMS mainly used a computer mouse for the pointing operator. Later, pointing was extended to accept hand gestures in touch screen interfaces [5, 6]. With the advancement of eye-tracking technology, more recent studies turned their focus on using eye-gaze for the pointing operator as it is faster and more convenient [7].

### 2.2  Video-based Online Evaluation

Human subject experiments are usually conducted in a laboratory with direct contact between the participants and experimenters. However, there are several studies that used online experiments in certain situations. The main advantages of conducting an experiment in the online format include having access to a large and diverse population of participants while reducing the cost and the time of the recruitment. [8, 9] verified the validity of remote data collection showing that the results are consistent with the lab-based experiments. Other studies also demonstrated that there is no significant difference between online and lab-based experiments [10, 11]. [12] discussed that the results of the online experiments are reliable but in some cases, depending on the type of the task, there are minor differences between these results and those of lab-based experiments. [13] compared several methods of conducting experiments, including video-based evaluation. It has been noted that the lack of supervision in online experiments may affect the results to some extent. While online or remote experiments have been used and studied in several areas, the efficiency of this research format is less known in the area of extended reality [14].

## 3  METHODOLOGY

In this study, three XR applications were developed using Unity and Mixed Reality Toolkit. The applications were implemented using built-in eye-tracking and hand-tracking features of Microsoft HoloLens 2. The XR environment includes five numerically labeled cubes. The user placed them in order of the numbers from bottom to top using three interaction modes including Eye-gaze & Pinch, Eye-gaze & Voice, and Drag & Drop.

## 3.1 GOMS-based Task Analysis

The GOMS model has been successfully used as a predictive tool for measuring task execution time especially when it is not feasible to evaluate an interface with human subjects. In this study, the task execution times of the three interaction modes were predicted by defining the operators for each mode and considering the goal of placing a target object on a target place as quickly and accurately as possible.

The types of operators vary depending on the user interface. Some of them are unique for each interaction mode and others are shared between the three modes. We assigned several operators to each mode and derived their values from the literature. For our task, Perceptual operators include Scanning (S) requiring 13 ms [15] and Pointing with Eye ($P_e$) requiring 230 ms [5]. Cognitive operators include Mentally Prepare (M) requiring 1350 ms [16] and Pause before the Speech (Pa) requiring 700 ms [17]. Motor operators include Hand Preparation (Pr) requiring 452 ms, Pointing with Hand ($P_h$) requiring 1046 ms, Grab with Hand ($G_H$) requiring 586 ms all adopted from [16], Grab and Release with Voice Command ($G_V$) requiring 130 ms each [17], Move with Hand (MV) requiring 700 ms [4], Release with Hand ($R_H$) requiring 520 ms [16], Adjusting the Accuracy (A) was not defined in the literature and it is task dependent. Therefore, we consider it as a parameter, and Hand Retraction (Re) requiring 746 ms [16]. Finally, the General operator includes System Waiting Time (W) that varies depending on the response time of the system. We measured this time duration for our task and it was 550 ms.

## 3.2 Remote Survey Evaluation

An online experiment using a Qualtrics form was conducted to evaluate three modes of interaction in an XR environment. Figure 1 shows videos of the interaction modalities (Figure 1 – (a), (b), (c)) along with a sample of the survey questions (Figure 1- (d)).

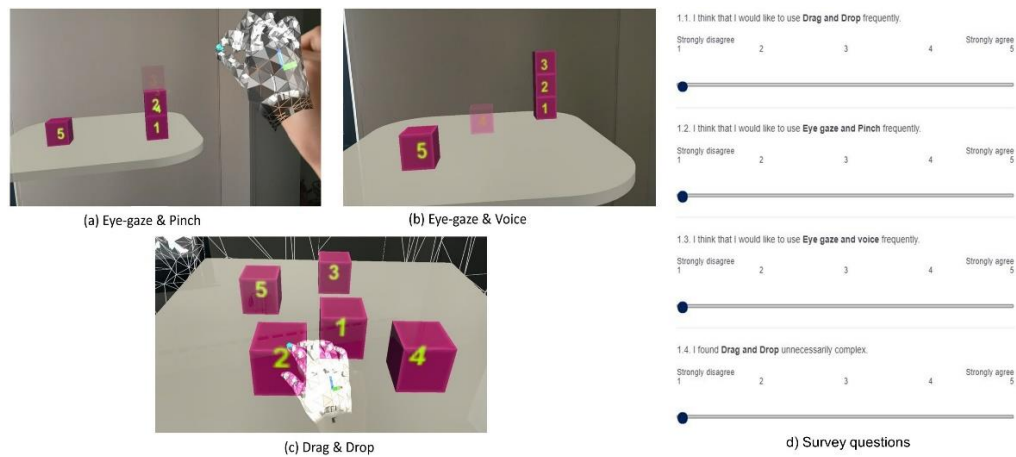

*Figure 1. Survey videos and questions*

Considering the substantial performance of gaze-based interactions demonstrated by previous research, we hypothesize that:

*H1:* Participants prefer multimodal gaze-based interactions over unimodal hand-based interactions in terms of usability.
*H2:* Participants prefer multimodal gaze-based interactions over unimodal hand-based interactions in terms of workload.

A total of 118 participants (82 males and 36 females) were qualified for testing the hypotheses. The age of the participants ranged from 18 to 65 with a mean of 25 (SD = 7.5). The three videos were shown to the participants. After watching the videos in the Qualtrics form, they were asked to answer questions from two standard questionnaires, System Usability Scale (SUS) and the National Aeronautics and Space Administration-Task Load Index (NASA-TLX), to evaluate the usability of the proposed interactions and their task load, respectively. We explicitly instructed them in the survey form to imagine that they are performing the tasks by themselves (rather than just watching the videos) and answer the questions based on their perception.

The SUS questionnaire consisted of 10 questions on a scale of 1 to 5 mainly about the usability of each mode (total of 30) to measure the effectiveness of the proposed interactions from several aspects. The NASA-TLX questionnaire consisted of 6 questions on a scale of 0 to 100 about the workload of the task for each mode (a total of 18) to subjectively measure the task load of the interactions. At the end of the questionnaires, we prepared 5 optional open-ended questions mostly about the potential applications and weaknesses of the proposed interactions (total of 15), and asked for the participants' ideas and feedback.

## 4 RESULTS

In this section, we present the findings of our experiments. First, we discuss the results of GOMS-based task analysis. We then explain the results of the survey questionnaires and participants' feedback on the three interaction modes.

## 4.1 GOMS-based Task Analysis Results

For model-based evaluation, we measured the total task execution time of the decomposed operators in three modes. Although the actual task involves five pick-and-place actions, only one pick-and-place action was analyzed,

- Eye-gaze & Pinch consists of 8 operators including $S + 2M + Pr + 2P_e + G_H + R_H = 3885$ (ms).
- Eye-gaze & Voice consists of 11 operators including $S + 2M + 2P_e + 2Pa + G_V + 2W + R_V = 5217$ (ms).
- Drag & Drop consists of 11 operators including $S + 2M + Pr + 2P_h + G_H + MV + A + R_H + Re = 6963 + A$ (ms) = minimum of 6963 (ms).

The results of the GOMS-based task analysis for the three interaction modes showed that the Eye-gaze & Pinch has the least task execution time equal to 4631 ms. The Eye-gaze & Voice came second with 5217 ms and Drag & Drop had the longest task execution time equal to 6963 ms.

## 4.2 Remote Survey Evaluation Results

The results of the statistical analysis for the SUS questionnaire showed that Drag & Drop outperformed the gaze-based modes and there was no significant difference between the two gaze-based methods. The results of the statistical analysis for the NASA-TLX questionnaire showed that the Eye-gaze & Voice was less demanding compared to the other two modes in terms of physical demand and there was no significant difference between Eye-gaze & Pinch and Drag & Drop. In the other sub-scales, Drag & Drop outperformed the gaze-based modes and there was no significant difference between the two gaze-based methods. Finally, based on the responses given to the open-ended questions, the gaze-based modes are useful for certain situations. Especially Eye-gaze & Voice is a better alternative for people with disabilities who are not able to use their hand or in dual-tasks where the user's hand is occupied with another task.

## 5 CONCLUSION, LIMITATIONS, AND FUTURE WORK

This study used model-based and video-based remote evaluations to assess multimodal gaze interaction modalities in XR. The results of the model-based evaluation showed that the gaze-based interactions offer shorter task completion time compared to the hand-based interaction. However, the results of the survey questionnaires suggested that the participants prefer the hand-based interaction over the gaze-based interactions in most aspects of the usability of the system and perceived workload.

The participants of this study were selected regardless of their experience and familiarity with XR to include a diverse range of viewpoints on the usability and workload aspects of the proposed interactions. Using remote video-based evaluation format provided several benefits alleviating some of the common shortcomings of in-lab experiments. First, it enabled us to access a large and diverse pool of participants. Second, due to the remote nature of the evaluations, there was no geographic limitations and scheduling concerns when recruiting participants. Third, since the tasks were already done and recorded, the whole process of data collection was significantly faster than that of in-lab experiments. Fourth, the participants were allowed to fill out the survey with their own pace unlike lab-based experiments where they might feel pressure from the experimenter or the environment. Finally, it minimized the risk associated with direct contact for both the experimenter and the participant particularly during the Covid-19 pandemic. However, compared to in-lab experiments, the remote format may pose some limitations. For example, the provided answers to survey questions may not accurately reflect the true utility of the tested interactions specifically in terms of workload and ease of use. Also, the participants may feel less committed without the experimenter's supervision during the experiment and thus, the results may vary to some extent compared to those of the lab-based experiments. Finally, due to the distractions associated with unsupervised experiments and uncontrolled environments, the participants may not be able to fully concentrate on the task and provide accurate responses.

For future studies, we aim to extend this work by performing in-person experiments. We intend to leverage additional performance measures such as the empirical task completion time and accuracy to evaluate the results. We also plan to measure the electrical activity of the brain while performing the task in each interaction mode. After conducting the in-person experiment, we will also be able to compare the findings of this study with those of the in-lab experiment to investigate the difference in the user responses based on their experience and their perception about the system.

**REFERENCES**
[1] David Kieras. 2009. Model-based evaluation. The Human-Computer Interaction: Development Process.
[2] Stuart K. Card. 1983. The Psychology of Human-Computer Interaction. Hillsdale, NJ: L. Eribaum Associates Inc.,
[3] Siti S. Azmi, Rajah J. R. Yusof, Thiam K. Chiew, Jadeera C. P. Geok, Gavin Sim. 2019. Gesture interfacing for people with disability of the arm, shoulder and hand (dash) for smart door control: GOMS analysis. Malaysian Journal of Computer Science.
[4] Henrik Tonn-Eichstädt. 2006. Measuring Website Usability for Visually Impaired People- A Modified GOMS Analysis. In Proceedings of the 8th International ACM SIGACCESS Conference on Computers and Accessibility
[5] Setthawong, P. and Setthawong, R., 2019. Updated Goals Operators Methods and Selection Rules (GOMS) with Touch Screen Operations for Quantitative Analysis of User Interfaces. International Journal of Adv. Sci. Eng. Inf. Technol.
[6] Heejin Jeong and Yili Liu. 2019. Computational Modeling of Touchscreen Drag Gestures using a Cognitive Architecture and Motion Tracking. International Journal of Human–Computer Interaction.
[7] Allsion Popola. 2011. The Effects of Eye Gaze based Control on Operator Performance in Monitoring Multiple Displays. Master Thesis. Embry-Riddle Aeronautical University - Daytona Beach, Human Factors and Systems Department.
[8] John H. Krantz and Reeshad Dalal. 2000. Validity of Web-based Psychological Research. In Psychological Experiments on the Internet. Academic Press.
[9] Paul Meyerson and Warren W. Tryon. 2003. Validating Internet Research: A Test of the Psychometric Equivalence of Internet and In-person Samples. Behavior Research Methods, Instruments, & Computers.
[10] Sandy J.J. Gould, Anna L. Cox, Duncan P. Brumby, and Sarah Wiseman. 2013. Assessing the Viability of Online Interruption Studies, In Proceeding of AAAI Conference on Human Computation and Crowdsourcing.
[11] Giuseppe Riva, Tiziana Teruzzi, and Luigi Anolli. 2003. The Use of the Internet in Psychological Research: Comparison of Online and Offline Questionnaires, CyberPsychology & Behavior.
[12] Frederic Dandurand, Thomas R. Shultz, and Kristine H. Onishi. 2008. Comparing Online and Lab Methods in a Problem-Solving Experiment. Behavior Research Methods.
[13] Alexandra Voit, Sven Mayer, Valentin Schwind, and Niels Henze. 2019. Online, VR, AR, Lab, and In-Situ: Comparison of Research Methods to Evaluate Smart Artifacts. In Proceedings of the 2019 CHI Conference on Human Factors in Computing Systems.
[14] Jack Ratcliffe, Francesco Soave, Nick Bryan-Kinns, Laurissa Tokarchuk, and Ildar Farkhatdinov. 2021. Extended Reality (XR) remote research: a survey of drawbacks and opportunities. arXiv preprint arXiv:2101.08046.
[15] Mary C. Potter, Brad Wyble, Carl E. Hagmann, and Emily S. McCourt. 2014. Detecting Meaning in RSVP at 13 ms per picture. Attention, Perception, & Psychophysics.
[16] Orlando Erazo and Jose Pino. 2015. Predicting Task Execution Time on Natural User Interfaces based on Touchless Hand Gestures. In Proceedings of the 20th International Conference on Intelligent User Interfaces,
[17] Joseph Jaffe, Beatrice Beebe, Stanley Feldstein, Cynthia Crown, and Michael Jasnow. 1970. Rhythms of Dialogue. Academic Press.